\pdfoutput=1

\documentclass[twocolumn, showkeys]{aastex631}
\usepackage{epstopdf}
\usepackage{xcolor}
\usepackage{xspace}
\usepackage{graphicx}
\usepackage{amssymb, amsmath}
\usepackage{natbib}
\usepackage{float}
\usepackage{mathptmx}
\usepackage{verbatim}
\usepackage{textcomp}
\usepackage{latexsym}
\usepackage{multirow}
\usepackage{wasysym}
\usepackage{academicons}
\usepackage{gensymb}
\usepackage{longtable}
\usepackage{ifthenx}
\usepackage{comment}
\usepackage{silence}
\usepackage{nameref}
\usepackage{placeins}
\usepackage{hyperref}
\usepackage{graphicx}
\usepackage{soul} 

\shorttitle{Proper Motion of HST-1}
\shortauthors{Thimmappa et al.}

\graphicspath{{./}{figures/}}

\begin{document}

\title{{\it Hubble} Study of the Proper Motion of HST-1 in the Jet of M87}

\correspondingauthor{R.~Thimmappa}
\email{rameshan.thimmappa@villanova.edu}

\author[0000-0001-5122-8425]{Rameshan Thimmappa}
\affiliation{Villanova University, Department of Physics, Villanova, PA 19085, USA}

\author[0000-0002-8247-786X]{Joey Neilsen}
\affiliation{Villanova University, Department of Physics, Villanova, PA 19085, USA}

\author[0000-0001-6803-2138]{Daryl Haggard} 
\affiliation{McGill University, Montreal, QC, Canada}

\author[0000-0001-6923-1315]{Mike Nowak} 
\affiliation{Washington University in St. Louis, St. Louis, MO 63130, USA}

\author[0000-0002-7263-7540]{Łukasz Stawarz}
\affiliation{Astronomical Observatory of the Jagiellonian University, Orla 171, 30-244 Kraków, Poland}

\begin{abstract}

The radio galaxy M87 is well known for its jet, which features a series of bright knots observable from radio to X-ray wavelengths. The most famous of these, HST-1, exhibits superluminal motion, and our analysis of {\it Chandra} data \citep{Thimmappa24} reveals a correlation between the X-ray flux of HST-1 and its separation from the core. This correlation likely arises from moving shocks in the jet, allowing measurement of the internal structure of HST-1 in the X-ray band. To follow up on these results, we use observations from the {\it Hubble} Space Telescope Advanced Camera for Surveys HRC/WFC/SBC channel and the Wide Field Camera 3 (WFC3)'s UVIS to analyze the image and flux variability of HST-1. Our analysis includes 245 ACS and 120 WFC3 observations from 2002-2022, with a total exposure time of $\sim345$ ks. We study the brightness profile of the optical jet and measure the relative separation between the core and HST-1 for comparison to the X-ray. We find that the X-ray and the UV/optical emission could arise from physically distinct regions. The measured proper motion of the knot HST-1 is 1.04$\pm$0.04 c from 2002-2005 and around 2.1$\pm$0.05 c from 2005-2022. We discuss the acceleration of the jet and the flaring synchrotron emission from HST-1 from optical to X-rays. 

\end{abstract}

\section{Introduction} 
\label{sec:intro}

Relativistic jets in radio galaxies carry large kinetic powers 
up to even $10^{47}$~erg~s$^{-1}$ and can travel up to 100 kpc from the core \citep{Blandford74, Bridle84, Blandford19}. These prominent jets can be seen in bright sources like Cen\,A \citep{Schreier81}, 3C 273 \citep{Harris87}, 3C\,120 \citep{Hjorth95}, Pictor\,A \citep{Wilson01}, 3C\,371 and PKS 2201+044 \citep{Sambruna07}, OJ\,287 \citep{Marscher11} and M87 \citep{Schmidt78}. These jets typically consist of a series of prominent knots. Some jet knots are interpreted as shocks produced either by collisions of inhomogeneities within the outflow \citep[the ``internal shock'' model; e.g.,][]{Rees78}, by interactions with the confining ambient medium \citep[the ``reconfinement shock'' model; e.g.,][]{Komissarov97}, or by interactions of jets with gas clouds or stars \citep[e.g.,][]{Wykes15}. 

The positions of the knots can be in good agreement between radio and X-ray observations \citep[e.g.,][]{Massaro15}, but in some other cases, there is a systematic offset 
between them, with the X-ray maxima typically located upstream of the corresponding radio knots \citep{Hardcastle03, Kataoka08, Reddy23}. \cite{Kadler04} found a good agreement between the radio and optical observations of the jet structure in the low-luminosity source NGC 1052. Jet knots have been extensively studied using \textit{Chandra} and \textit{Hubble} observations \citep[e.g., PKS~1136$-$135;][]{Cara13}; see also the detailed study of 4C+19.44 by \citet{Harris17}.

The radio galaxy M87 ($z$ = 0.0043) is one of the nearest active galaxies and hosts a supermassive black hole \citep[$\sim6.5\times10^{9}M_{\odot}$,][]{EHT19}. Its relativistic jet originates from the black hole with an estimated kinetic power of $\sim10^{44}$~erg~s$^{-1}$ \citep{Stawarz06, Prieto16}, and exhibits a series of bright knots up to a few kpc distances. This relativistic jet is extensively studied in all bands, from radio to X-rays \citep{Owen89, Biretta95, Sparks96, Biretta99, Marshall02, Perlman05, Perlman11, Meyer13, EHT21, Cui23, Algaba24, Roder25}. For example, a proper motion study of the jet in M87 was performed using Very Long Baseline Interferometry \citep[VLBI;][]{Reid89, Cheung07, Giroletti12, Walker18}, the Very Large Array \citep[VLA;][]{Biretta95}, and the {\it Chandra} X-ray Observatory \citep{Snios19, Thimmappa24}.

One of the most exciting features of the jet of M87 is HST-1, a prominent knot located $\simeq$70 pc (projected) from the core, as seen by the {\it Hubble} Space Telescope \citep{Biretta95, Biretta99}. HST-1 exhibited a bright radio to $\gamma$-ray flare from 2000-2008, peaking in 2005 \citep{Harris03, Aharonian06, Ly07, Cheung07, Madrid09, Harris09, Acciari10,  Rieger12, Yang19}. It has been the subject of extensive study due to its unusual properties, including superluminal motion, high-energy emission, and significant structural changes. HST-1 has shown extreme variability in X-ray and optical, with increases by factors of 50 and 4, respectively, over short periods \citep{Harris06}.
 
In this paper, we study the proper motion and internal structures (i.e., sub-components) of HST-1 using Hubble's Advanced Camera for Surveys (ACS) and Wide Field Camera 3 (WFC3). With its excellent optical imaging capabilities, {\it Hubble} can resolve structures in the jet and track the movement and evolution of these flares. In our previous paper \citep{Thimmappa24}, we found a correlation between the flux of HST-1 and its offset from the core using {\it Chandra} observations from 2002 to 2008. A list of Chandra datasets, obtained by the Chandra X-ray Observatory, contained in~\dataset[DOI: https://doi.org/10.25574/cdc.203]{https://doi.org/https://doi.org/10.25574/cdc.203}. We developed a toy model for the correlation to show unresolved additional emission regions. Here, we apply the same model for the same period of {\it Hubble} data to study the jet structure in the optical, which provides complementary information to X-ray observations and helps to build a complete picture of the jet's behavior. In addition, we investigate the proper motion of HST-1 from 2002 to 2005 and from 2005 to 2022. This paper aims to understand the superluminal motion and structural changes of HST-1 in the jet of M87 and the relationship between X-ray and optical emission. 

This paper is organized as follows. In section \ref{sec:data_analysis}, we provide the details of the {\it Hubble} data. In Section \ref{sec:results}, we present the analysis of the jet and our model for the flux-offset correlation of HST-1. We present joint fitting of X-ray and optical observations, compare our results to previous measurements of the proper motion of HST-1, and discuss the implications in Section \ref{sec:discussion}.

\begin{figure}[ht!]
    \centering 
    \includegraphics[width=\columnwidth]{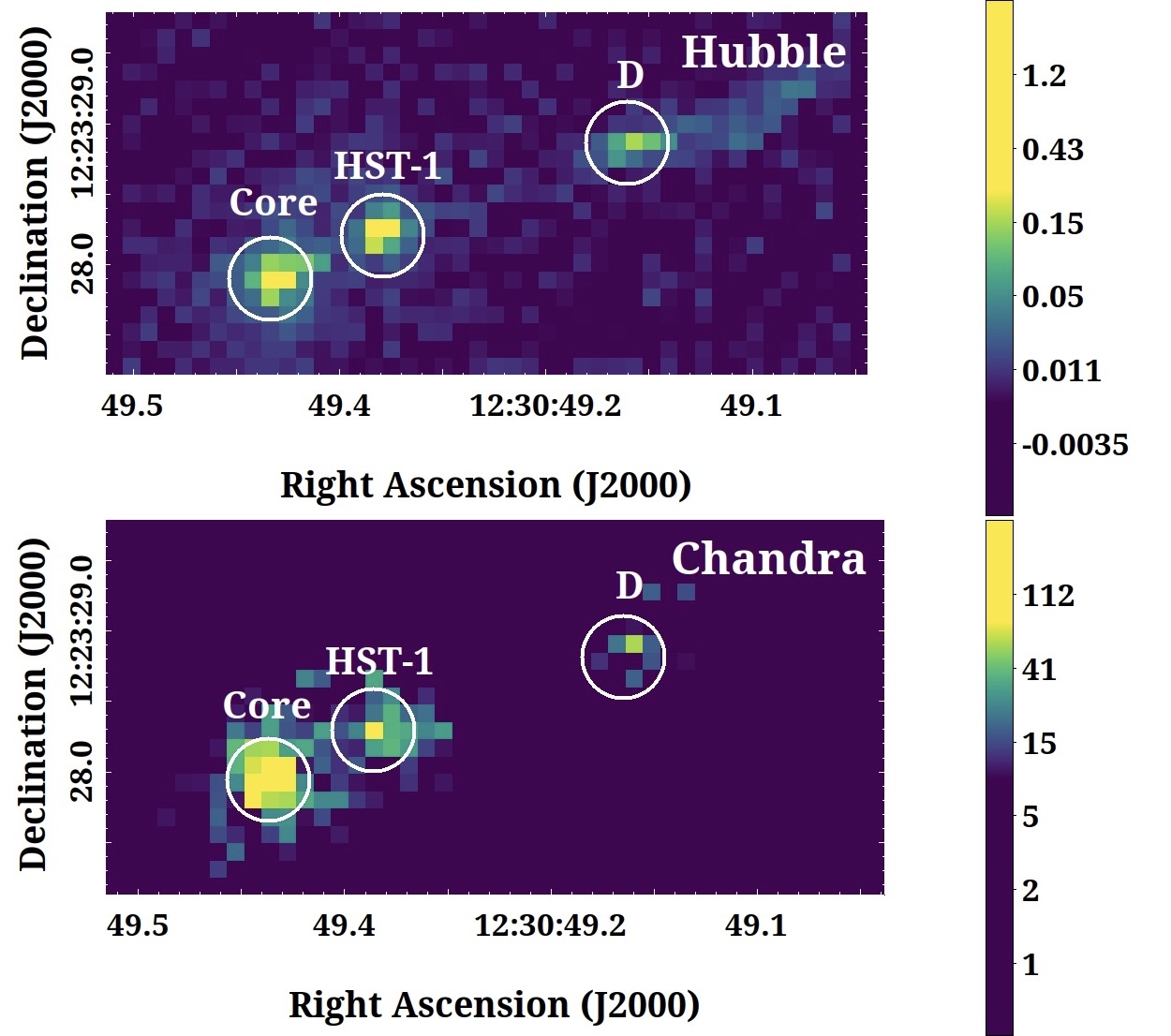}
    \caption{\textit{Top:} Drizzled \textit{HST}/ACS-HRC image of the M87 jet (Program ID J8L001010, F220W filter). 
    The color bar is flux density in units of erg\,s$^{-1}$\,cm$^{-2}$\,\AA$^{-1}$\,pixel$^{-1}$. 
    \textit{Bottom:} Exposure-corrected, deconvolved \textit{Chandra}/ACIS image of ObsID~3975 at 0.25-pixel resolution, with the color bar is flux in photons\,cm$^{-2}$\,s$^{-1}$\,pixel$^{-1}$.}
    \label{fig:jet}
\end{figure}

\section{{\it Hubble} Data Analysis} 
\label{sec:data_analysis}

For this study, we use 245 ACS \citep{Ford98} pointings from 2002 to 2016 with a total exposure time of $\sim91$ ks and 120 WFC3 pointings from 2012 to 2022 with a total exposure time of $\sim$ 254 ks. The ACS observations include High-Resolution Channel (HRC, $\sim$0$''$.025/pixel), Wide Field Channel (WFC, $\sim$0$''$.05/pixel), and Solar Blind Channel (SBC, $\sim$0$''$.032/pixel) CCD instruments with F220W, F250W, F330W, F475W, F606W and F814W filters (from near-UV, visible to the near-IR band). These data contain 199 HRC ObsIDs, 44 WFC ObsIDs, and 2 SBC ObsIDs. The WFC3 observations include all UVIS channels ($\sim$ 0$''$.04/pixel) with F225W, F275W, F390W, F475W, F606W and F814W filters. All of the data presented in this article were obtained from the Mikulski Archive for Space Telescopes (MAST) at the Space Telescope Science Institute \footnote{{\url{https://mast.stsci.edu/search/ui/\#/hst}}}. The specific ObsIDs analyzed can be accessed via \dataset[doi:10.17909/a7dy-5050]{https://doi.org/10.17909/a7dy-5050}, and we also used the subproducts under the association ObsID.

For our analysis, we use the {\fontfamily{qcr}\selectfont AstroConda} and {\fontfamily{qcr}\selectfont DrizzlePac\footnote{{\url{https://www.stsci.edu/scientific-community/software/drizzlepac}}}'s} \citep{Avila15} {\fontfamily{qcr}\selectfont AstroDrizzle} \citep{Gonzaga12, Hoffmann21} software packages, which are maintained by the Space Telescope Science Institute \citep{STSCI12}. The raw files are processed on the association table files using the ACS and WFC3 calibration pipeline {\fontfamily{qcr}\selectfont calacs} to produce flat-field exposure images. The pipeline subtracts the bias and dark current and applies the flat field to the raw CCD data \citep{Sirianni05}. These images are corrected for charge transfer inefﬁciency using the current ACS and WFC3 pipeline \citep{Anderson18}. 
The {\fontfamily{qcr}\selectfont updatewcs} tool updates the header keywords with new WCS information and distortion corrections in the flat field individual file. The {\fontfamily{qcr}\selectfont TweakReg} software aligns a set of images for better accuracy ($<$0.1 pixels). The given reference image's WCS aligns all input images. Then, we use {\fontfamily{qcr}\selectfont AstroDrizzle} to remove geometric distortion, detector artifacts, and cosmic rays. The {\fontfamily{qcr}\selectfont AstroDrizzle} can produce a combined single image of {\fontfamily{qcr}\selectfont drz} ﬁts ﬁle that incorporates all of the individual exposures, which are electrons/s, and is used for final image analysis. The drizzled image of the F220W filter is presented in Figure \ref{fig:jet} at the top.

\begin{figure*}[ht!]
	\centering 
    \includegraphics[width=0.98\textwidth]{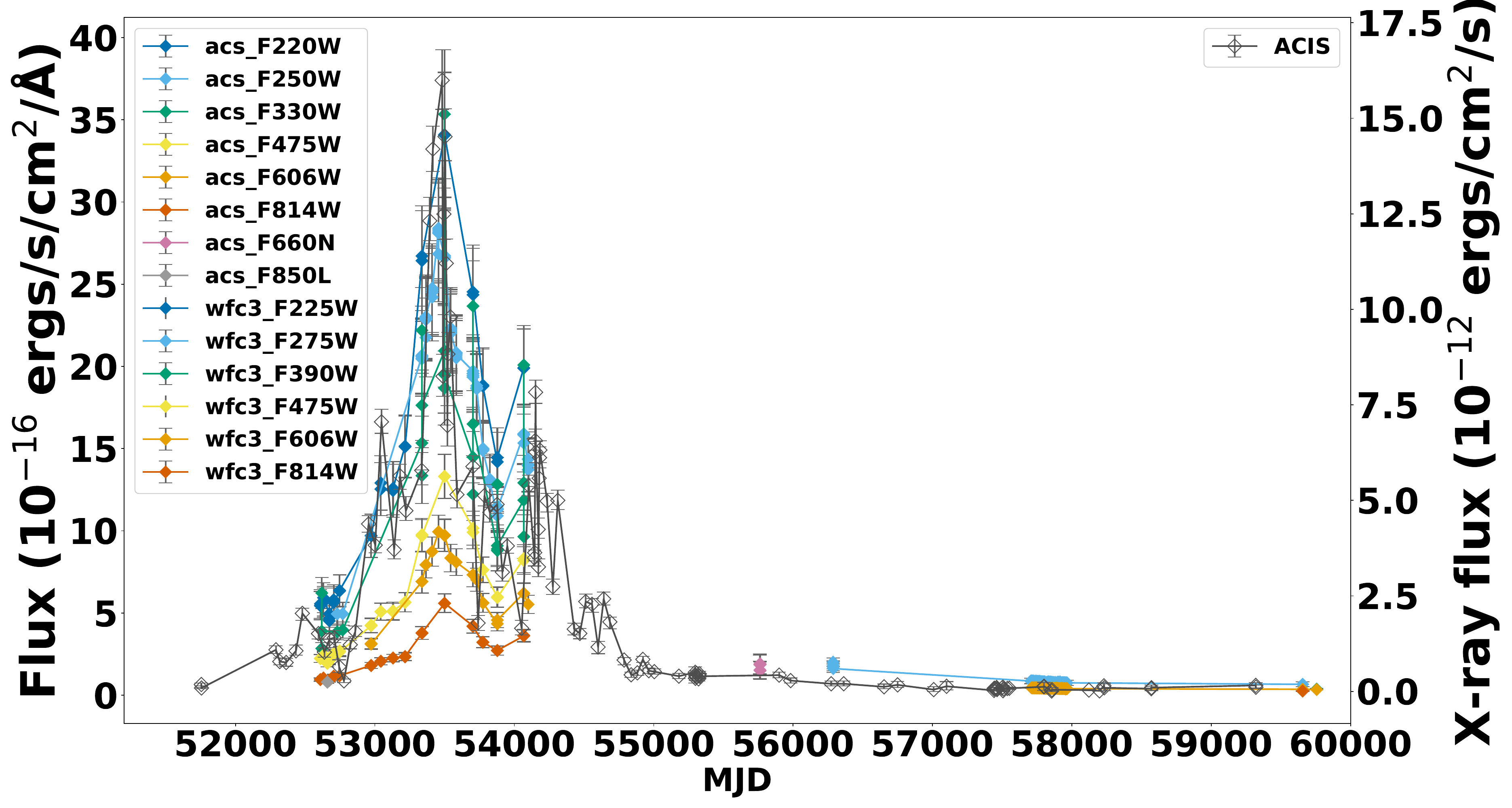}
	\caption{The {\it Chandra}/{\it Hubble} lightcurves of HST-1. The flux rises around 2005 for both observations and drops in 2008, then remains the same until 2022. Blue color shows {\it Chandra} flux, and other color lines show {\it Hubble} filters. The complete dataset is available in machine-readable format with accompanying metadata.}
     \label{fig:flux_ACS_vs_ACIS}
	\centering 
        \includegraphics[width=0.98\textwidth]{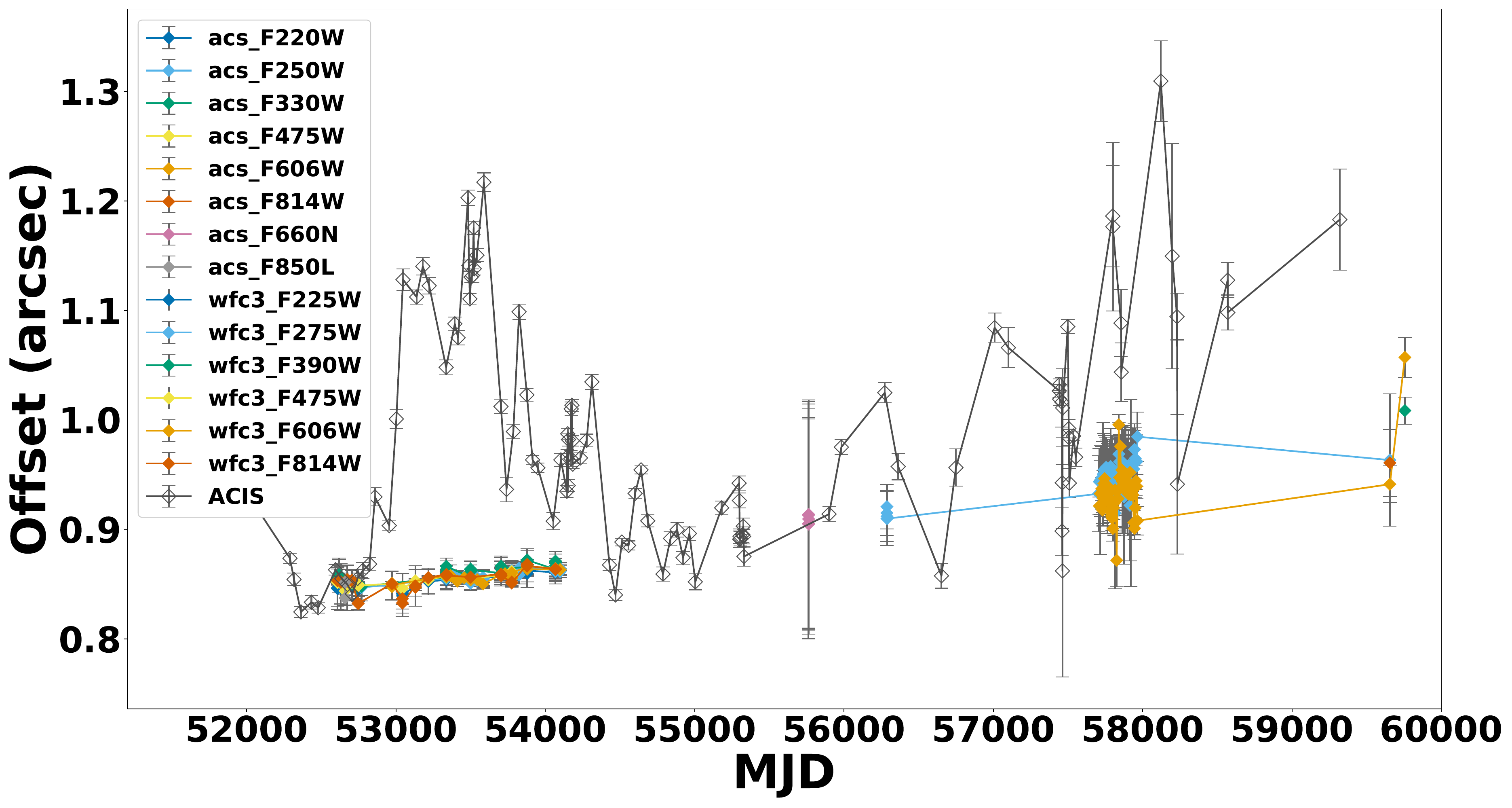}
	\caption{The offset of HST-1 from the core of M87 as seen by {\it Chandra} and {\it Hubble} from 2002 to 2022.}
     \label{fig:Offset_ACS_vs_ACIS}
\end{figure*}

\section{Results}
\label{sec:results}

\subsection{The Lightcurve of HST-1}
\label{sec:lightcurve}

We produce drizzled images to study the proper motion of HST-1 and its flux-offset correlation in the jet of M87. The centroid position of the core and HST-1 are selected with $0''.25$ radius using SAOImage DS9\footnote{\url{https://ds9.si.edu}} \citep{Joye03}, as shown in Figure \ref{fig:jet} at the top. We measure the flux of HST-1 for each observation from ACS/HRC, WFC, SBC, and WFC3/UVIS data. The flux and its uncertainties are calculated from the following equations:

\begin{equation}
    F = \text{Net counts} \times \text{PHOTFLAM}
\label{eqn:model_Flux}
\end{equation}
\begin{equation}
    \sigma_F = F \cdot \sqrt{
        \left( \frac{\sigma_{\text{counts}}}{\text{NetCounts}} \right)^2 +
        \left( \frac{\sigma_{\text{PHOTFLAM}}}{\text{PHOTFLAM}} \right)^2
    }
    \label{eqn:model_Flux_Error}
\end{equation}

\noindent where the net counts are the total count rate from the source (in e\textsuperscript{$-$}/s), and {\fontfamily{qcr}\selectfont PHOTFLAM} is a scaling factor that converts an instrumental flux (measured in electrons per second) to a physical flux density.  
We extract the net counts within a radius of 0$''$.25 using the {\fontfamily{qcr}\selectfont dmstat} tool in CIAOv4.17 \citep{Fruscione06} analysis threads\footnote{\url{http://cxc.harvard.edu/ciao/threads/}}. $\sigma_{\text{counts}}$ is the uncertainty of the netcounts, which is the background-subtracted total counts from a source region. The uncertainty of {\fontfamily{qcr}\selectfont PHOTFLAM} is about 1\%, as the calibration provides a photometric internal precision of about 1\% for all filters \citep{Bohlin16, Newman24}. The resulting light curve of HST-1 is presented in Figure \ref{fig:flux_ACS_vs_ACIS}. Each line represents a different ACS and WFC3 filter.

From Figure \ref{fig:flux_ACS_vs_ACIS}, all the ACS filter fluxes increase during flaring activity from 2002 to 2005 and then return to the baseline level in 2008. After 2008, the fluxes remain roughly the same. Overall, the flare behavior is similar across all wavelengths, although the flare amplitude is smaller at longer wavelengths. For example, the F814W flux increases from $\sim$1$\times$10$^{-16}$\,erg\,cm$^{-2}$\,s$^{-1}$\,\AA$^{-1}$ to $\sim$5$\times$10$^{-16}$\,erg\,cm$^{-2}$\,s$^{-1}$\,\AA$^{-1}$, while the F220W flux rises from $\sim$5$\times$10$^{-16}$\,erg\,cm$^{-2}$\,s$^{-1}$\,\AA$^{-1}$ to $\sim$35$\times$10$^{-16}$\,erg\,cm$^{-2}$\,s$^{-1}$\,\AA$^{-1}$. This corresponds to a flux ratio of about 5 for F814W and about 7 for F220W, and a flux increase of $\sim4\times10^{-16}$ \,erg\,cm$^{-2}$\,s$^{-1}$\,\AA$^{-1}$ and $\sim30\times10^{-16}$ \,erg\,cm$^{-2}$\,s$^{-1}$\,\AA$^{-1}$ for F814W and F220W, respectively.

After extracting the lightcurve, we select the centroid of the core and HST-1 regions in the image as shown in Figure \ref{fig:jet} top panel. For each ObsID, we fit the radial profile and measure the offset of HST-1 from the core using a one-dimensional Gaussian fitting with the {\fontfamily{qcr}\selectfont Sherpa}\footnote{\url{https://cxc.cfa.harvard.edu/sherpa}} package \citep{Freeman01}. The model fitting is performed with the \texttt{Levenberg-Marquardt} method and \texttt{chi2gehrels} statistic. The uncertainties of the parameters are estimated using {\ttfamily conf}, which computes confidence intervals for the given model parameters. The resulting offset of HST-1 is shown in Figure \ref{fig:Offset_ACS_vs_ACIS} and increases slowly in the {\it Hubble} data. For the comparison, the X-ray flux and offset of HST-1 are plotted along with the UV/optical results and are shown in Figure \ref{fig:flux_ACS_vs_ACIS} and Figure~\ref{fig:Offset_ACS_vs_ACIS}.

The X-ray offset of HST-1 is systematically larger and more variable than the optical offsets measured with {\it Hubble}. The measured X-ray offsets have a mean of $0''.984$, and a standard deviation of $0''.106$, compared to $0''.858$, and $0''.011$ for the optical data. These results suggest that the X-ray and optical emissions very likely originate from physically distinct regions within HST-1, possibly tracing different particle-acceleration processes at different working surfaces along the jet. 

To investigate the correlation between flux and offset of HST-1 during the flare activity (Figure~\ref{fig:flux_ACS_vs_ACIS}), we attempt to apply the toy model we previously used for the X-ray data in \cite{Thimmappa24} to each Hubble/ACS filter separately, and also for the X-ray and UV/optical data simultaneously, as discussed in Section \ref{sec:model}. However, the flux offset model did not provide a satisfactory description of the data. We therefore attempted an alternative method to measure the proper motion, as detailed in Section \ref{sec:proper_motion}.

\subsection{The Flux-Offset Model}
\label{sec:model} 

Figure~\ref{fig:Offset_ACS_vs_ACIS} shows that there is no clear correlation between the optical flux and its offset from the core. The measured mean offset of the X-ray emission, moreover, differs from the optical emission by $\sim0\farcs126$ relative to the baseline position of HST-1. Previously, when analyzing the Chandra data alone \citep{Thimmappa24}, we proposed a toy model to account for the observed X-ray flux--offset correlation of HST-1, assuming multiple emission regions within the knot. To characterize the optical data analyzed in this paper, we first applied the same flux–offset correlation model. Our model describes the offset $O(t)$ of HST-1 from the core by: 
\begin{equation}
    O(t) = D + C(t) , 
\label{eqn:offset_HST-1}
\end{equation}
where $D$ is the initial offset of HST-1 before the flare and $C(t)$ is the centroid position. The centroid position is equal to the ratio of the variable flux $f(t)$ and total flux $f(t)+F_0$ times the downstream separation $d$: 
\begin{equation}
 C(t) = \frac{f(t)}{f(t)+F_0} \times [d + V_{ws}(t-t_{o})].
 \label{eqn:centroid_HST-1}
\end{equation}
When $d$ is small, the flux-weighted centroid can be roughly approximated by the best-fit position of HST-1. The downstream region may be linked to a working surface in a jet, where ejecta collide and form internal shocks \citep{Cant00, Mendoza09}. $V_{\mathrm{ws}}$ is the propagation speed of the working surface (shock front), which is responsible for the centroid variations. This speed can be comparable to the bulk velocity inferred from proper motion measurements.

In our model for the lightcurve of HST-1, we approximate the exponential rise/decay lightcurve of a working surface as
\begin{equation}
f(t) = A e^{-(\frac{t-t_{o}}{\tau_{1}}+\frac{\tau_{2}}{t-t_{o}})},
\label{eqn:flux_HST-1}
\end{equation}
where $A$ is the amplitude, $t$ is the time, $\tau_{1}$ and $\tau_{2}$ are fall and rise times. 

\subsubsection{Applying the flux/offset model to {\it Hubble} data}
\label{sec:individual_fits}

In our initial attempt, we first separate all filters and then fit our toy model simultaneously to the flux and offset of HST-1, assuming a fixed working-surface velocity. For our model fitting, we converted all ACS filter fluxes from erg/s/cm$^{2}$/\AA\, to erg/cm$^{2}$/s by multiplying the bandwidth of each filter to match the X-ray flux units. We used a working surface velocity of 6.4\,c for X-ray data, which was previously reported by \cite{Snios19}, and which we used while fitting the flux/offset model to X-ray data \citep{Thimmappa24}. All common parameters of the flux and offset models are linked using the {\fontfamily{qcr}\selectfont Sherpa} package. We performed fits for the centroid model on all of the filters individually. 

\begin{figure}[ht!]
	\centering 
          \includegraphics[width=\columnwidth]{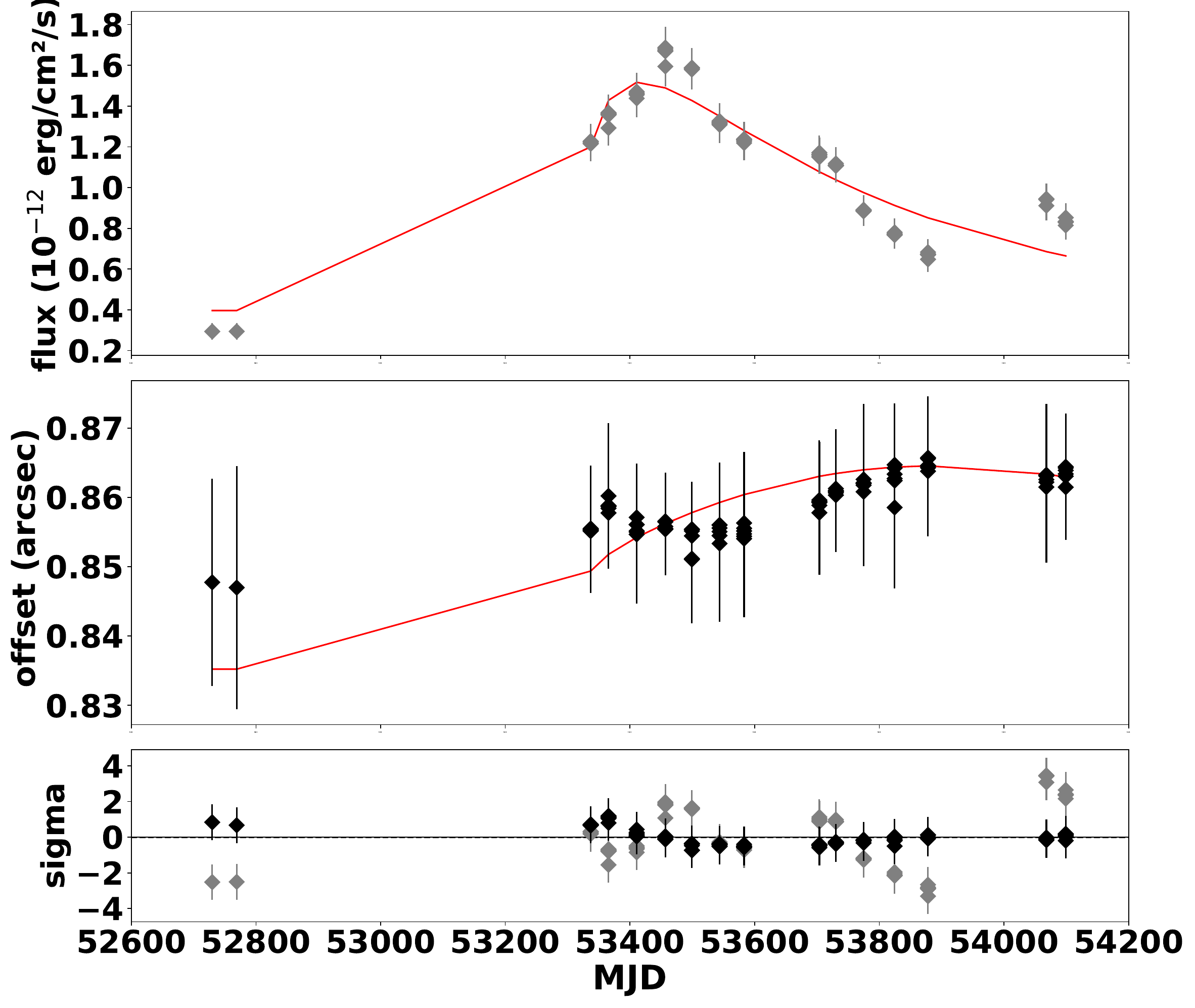}
	\caption{Joint fits of Model 1 (i.e., Equations \ref{eqn:flux_HST-1} and \ref{eqn:offset_HST-1}) to flux (gray) and offset (black) of HST-1 for the F250W filter.}
   \label{fig:250}
\end{figure}

As an example, we present the results of the toy-model fit to the flux–offset relation in the F250W filter. The fitting parameters are given with 90\% confidence errors as follows: $F_0$ is $0.40\pm0.02\times 10^{-12}$ erg/cm$^2$/s and the amplitude is $2.1\pm0.25\times 10^{-12}$ erg/cm$^2$/s with high uncertainty. The initial offset (D, baseline) is $0''.835\pm0''.006$. The distance ($d$) from the baseline is $\simeq$0$''.018\pm0''.009$. The start time is 53300$\pm$10 days, the fall time is 400$\pm$30 days, and the rise time is 40$\pm$10 days. The reduced $\chi^2$ value is 1.68, which is not a very good fit, but the best fit among all filters. The fitted plot of filter F250W is given in Figure \ref{fig:250}.  

The model provides a better fit to the UV data than the optical and IR bands, but the downstream distance is still d$<$0$''$.02 in all bands of {\it Hubble} data. For reference, the main flare separation from the baseline position of HST-1 is $d=0''.2486\pm0''.0074$ in X-rays. After fitting each filter separately, none provides strong evidence of additional emission regions well separated from the baseline position of HST-1. We found similar results when modeling all six filters jointly. The downstream distance $d\sim0$ suggests that the UV/optical emission is not co-spatial with the X-ray emission, which raises a question about the correlation between the two bands inside HST-1. 

We also perform a simultaneous fit for the X-ray and UV/optical data to determine whether the {\it Hubble} data is consistent with the additional emission regions observed by {\it Chandra}. However, our simultaneous fit with X-ray and optical emitting regions is inconsistent with the data, and the preferred locations for the optical emission are at d$\sim$0. A detailed discussion is in the Appendix~\ref{sec:Chandra_vs_Hubble}.

\subsection{Proper Motion of HST-1}
\label{sec:proper_motion}

Based on the Hubble observations and previous Chandra studies of the flux and offset of HST-1 during its flare activity (Figure~\ref{fig:flux_ACS_vs_ACIS} and ~\ref{fig:Offset_ACS_vs_ACIS}), we conclude that the X-ray and optical emissions originate from distinct locations within the HST-1 region. Moreover, the failure of our toy model, which successfully accounts for the Chandra data, when applied to the Hubble data implies that the optical flux is dominated by a single component, whereas the X-ray flux originates from multiple unresolved emission regions. On the other hand, there is a slight increase in the optical offset (Figure~\ref{fig:Offset_ACS_vs_ACIS}) from 2002 to 2022. Therefore, we use the corresponding offset values to measure the proper motion of HST-1 in the Hubble data over this period. Previous studies of high-resolution observations from the radio, optical, and X-ray bands have revealed the proper motions of the jet in M87, highlighting its complex kinematic structure. The inner jet shows subluminal motions with apparent velocities around 0.3 c, as detected using VLBI observations \citep{Reid89, Dodson06, Kovalev07}. HST-1 exhibits superluminal motion, and its position angle changes significantly \citep{Chen11b, Giroletti12, Asada14}. In contrast, the outer jet--particularly beyond HST-1--shows both subluminal and superluminal components, as observed by the VLA and the HST \citep{Biretta95, Meyer13}. Here, we focus on the motion of HST-1 itself.

\subsubsection{Proper Motion From 2002-2006}
\label{sec:velocity_calculation1}

\begin{figure}[ht!]
	\centering 
{ 	\includegraphics[width=0.49\columnwidth]{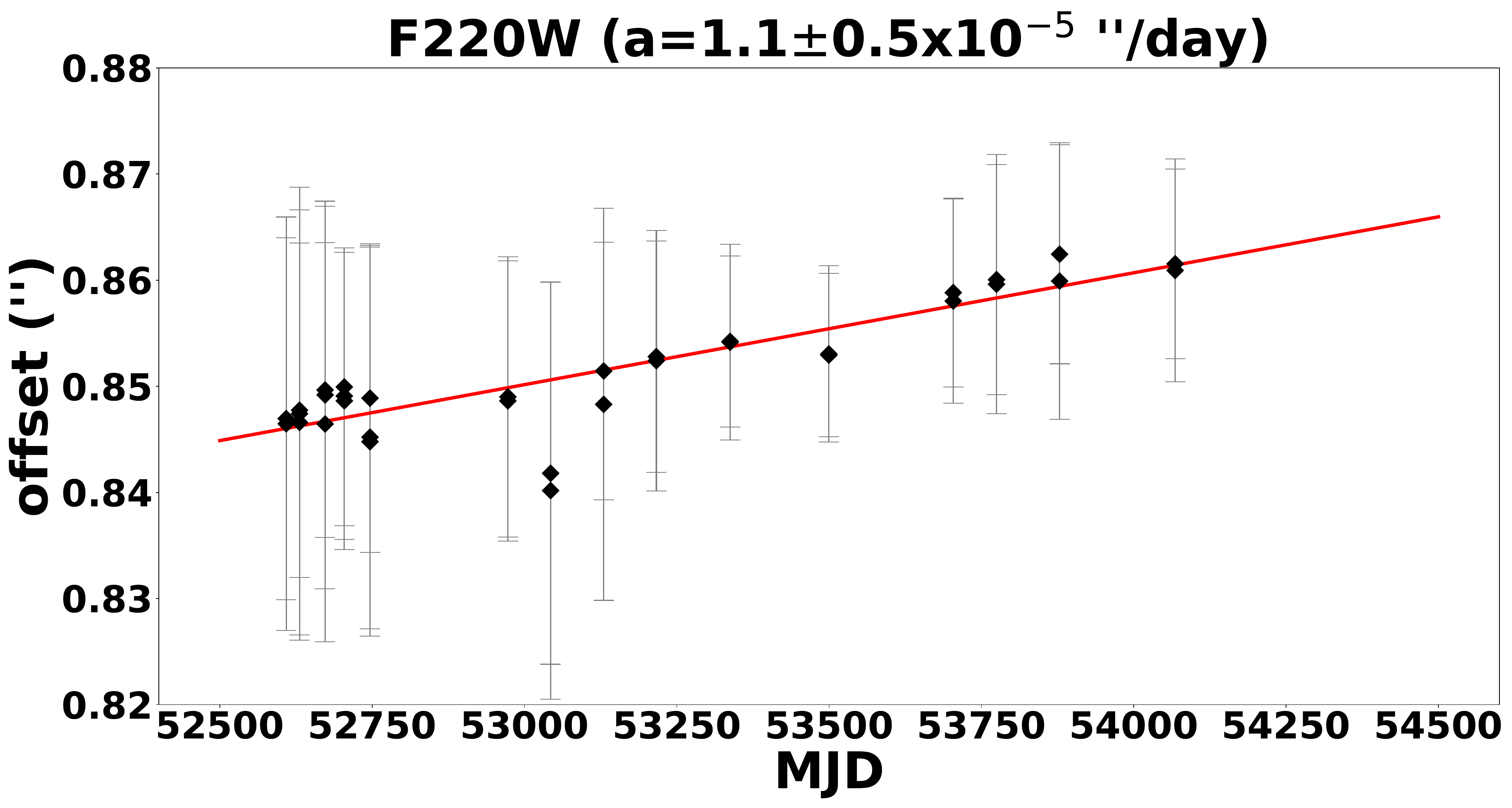}
 	\includegraphics[width=0.49\columnwidth]{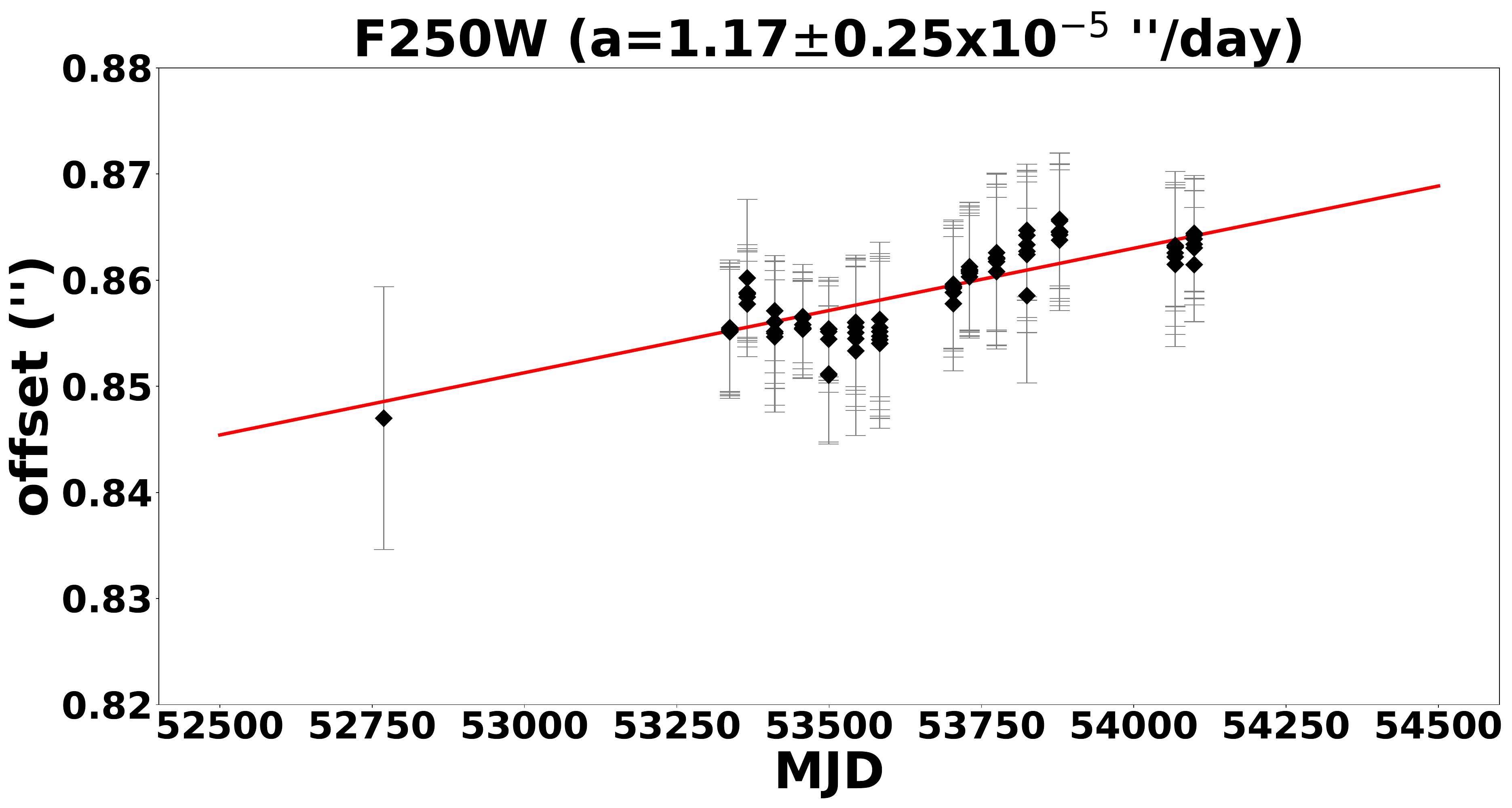}
    \includegraphics[width=0.49\columnwidth]{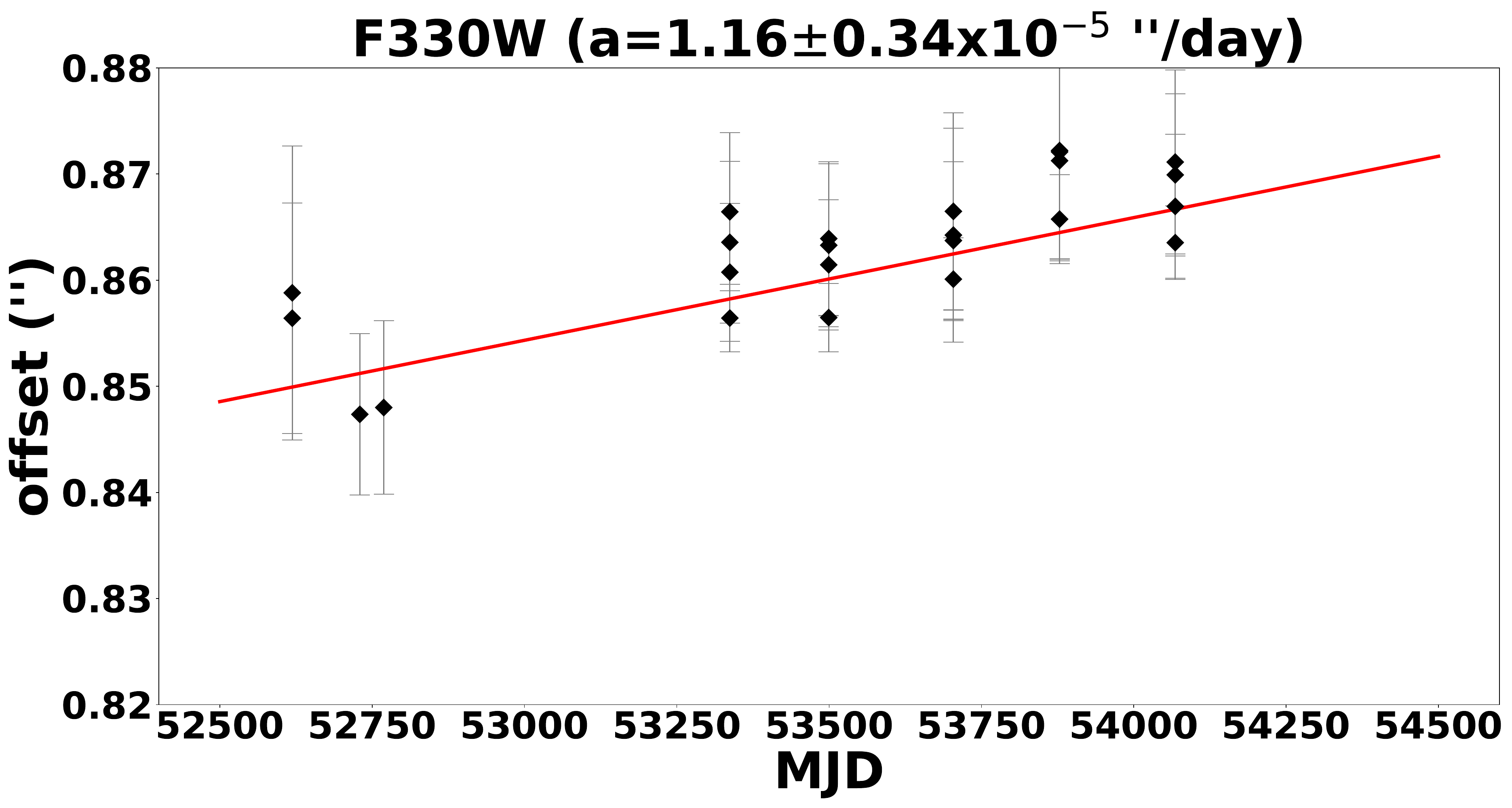}
 	\includegraphics[width=0.49\columnwidth]{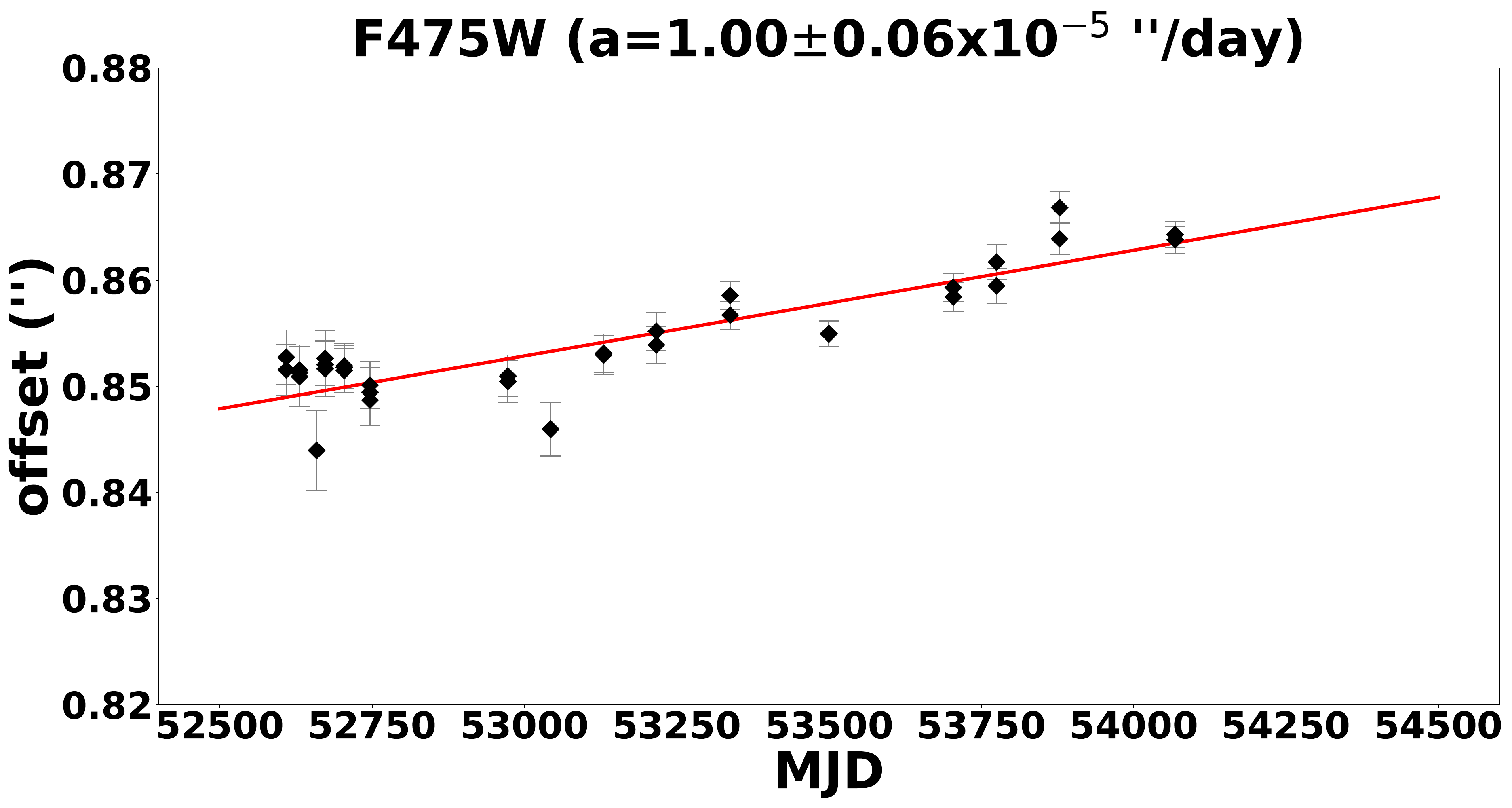}
    \includegraphics[width=0.49\columnwidth]{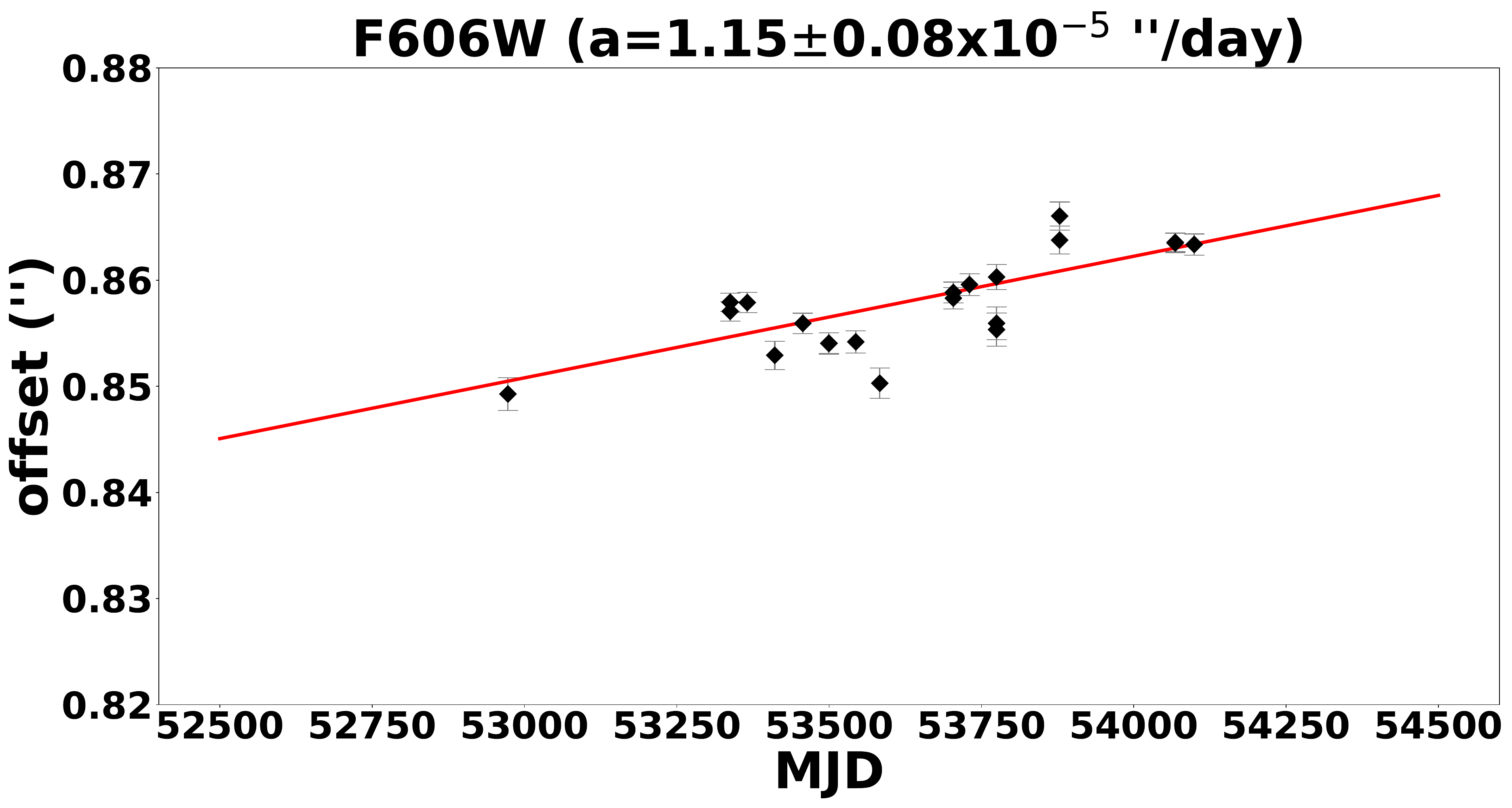}
    \includegraphics[width=0.49\columnwidth]{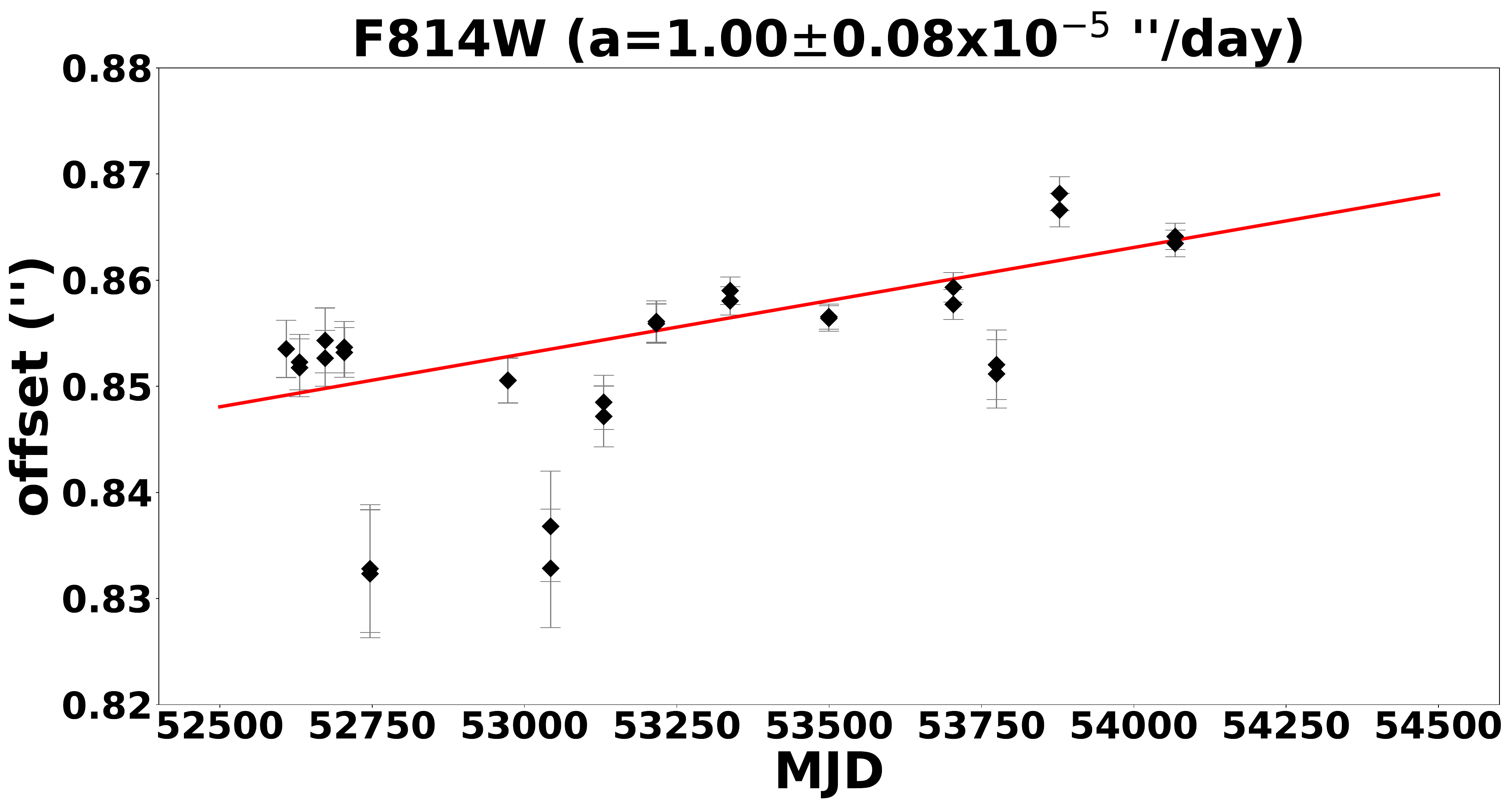} }
    \caption{Linear regression for the 2002–2006 offset vs time in different ACS filters. Left panel: F220W, F330W, and F606W. Right panel: F250W, F475W, and F814W. The model is shown as a red line.}
     \label{fig:Linear_regression_offset_ACS}
\end{figure}

For all ACS filters from 2002 to 2006, we observed no clear flux-offset correlation; in this case, the {\it Hubble} data can give us the speed of the jet knot. We perform a linear regression using {\fontfamily{qcr}\selectfont curve\_fit} in Python for the offset vs. MJD of HST-1, incorporating the measured uncertainties on the offset. This is an alternative to our toy model (see section \ref{sec:Chandra_vs_Hubble}). With this fit, we find an increase in the offset of HST-1, as shown in Figure \ref{fig:Linear_regression_offset_ACS}. The measured average speed of HST-1 is 0.99$\pm$0.04$\times$\,10$^{-5}$ $''$/day for ACS filters. This value corresponds to a bulk superluminal speed of 1.04$\pm$0.04 c for HST-1 during this early phase.

\begin{figure}[ht!]
	\centering 
     	\includegraphics[width=\columnwidth]{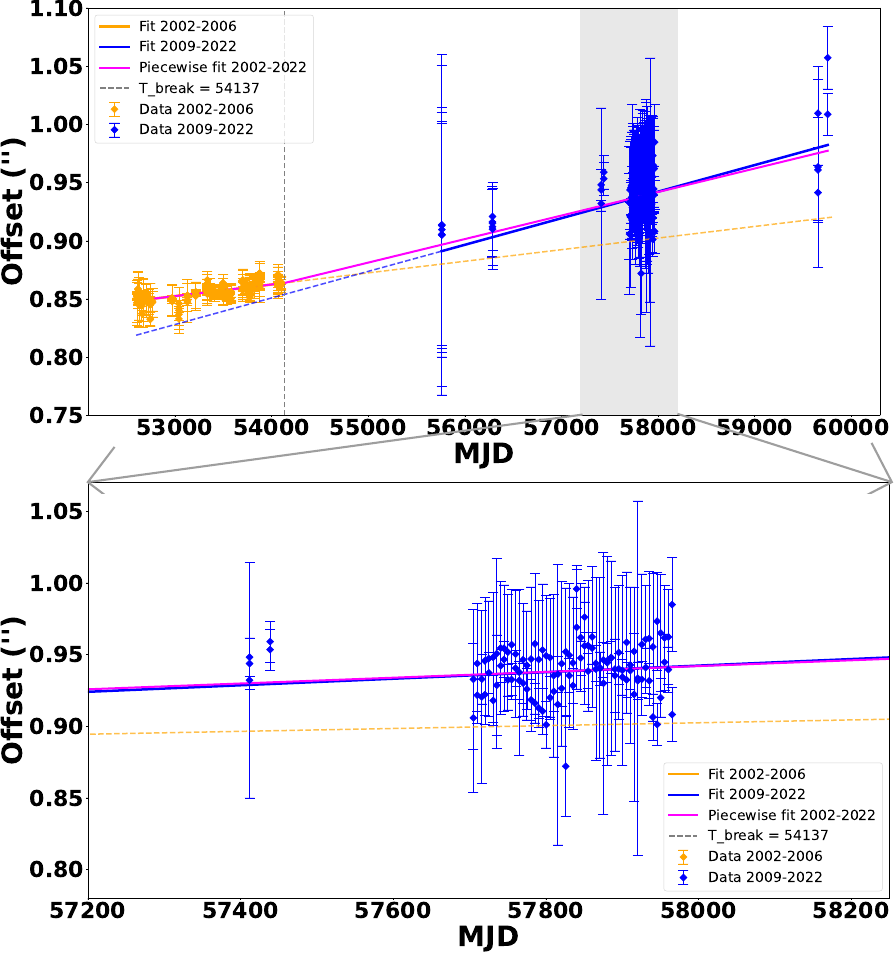}
    \caption{\textit{Top panel:} Piecewise regression model for the offset of HST-1 vs time during the period 2002-2022 (see Section \ref{sec:velocity_calculation3}). The model is shown in the magenta line, the orange line shows the pre-break model (solid before the break and dashed after the break), the blue line shows the post-break model (solid after the break and dashed before the break), and the gray line shows the break point at 54137 days. \textit{Bottom panel:} zoomed view of the same figure between MJD 57200 and MJD 58200.}
   \label{fig:speed_2002-2022}
\end{figure}

\subsubsection{Proper Motion From 2009-2022}
\label{sec:velocity_calculation2}

We continue to measure the proper motion of HST-1 after the flare from 2009 to 2022 using all {\it Hubble}/WFC3 data. This period shows continued motion of HST-1, as shown in Figure \ref{fig:speed_2002-2022} with an increase in velocity following the flare, indicated by the dotted blue line, which extrapolates the early-time slope to later times. Using the linear regression model, we measure a bulk speed of HST-1 of 2.3$\pm$0.5$\times$\,10$^{-5}$ $''$/day, which corresponds to a measured superluminal speed of 2.4$\pm$0.5 c. The velocities derived from UV/optical/IR data remain significantly lower than the apparent velocity from X-ray observations \citep[$\simeq6.6\pm0.9$\,c,][]{Thimmappa24}, suggesting that distinct emission regions may dominate at different wavelengths. The sub-components inside the HST-1 knot may contain different velocities for different positions at different wavelengths \citep[][Figure 12 therein]{Hada24}.  

\subsubsection{Proper Motion From 2002-2022}
\label{sec:velocity_calculation3}

In sections \ref{sec:velocity_calculation1} and \ref{sec:velocity_calculation2}, we measure HST-1's speed and find the changes between the two epochs. Therefore, to characterize the long-term evolution of HST-1's motion, we fit a linear regression model using a piecewise function for the data from 2002 to 2022. The best-fit model is shown in Figure \ref{fig:speed_2002-2022}, and is given as follows:
\begin{equation}
\text{Offset}(t) =
\begin{cases}
a + b_1(t - t_0), & \text{for } t < t_b \\
a + b_1(t_b - t_0) + b_2(t - t_b), & \text{for } t \geq t_b
\end{cases}
\label{eqn:piecewise}
\end{equation}
where $t_0$=53775 days is the start time of the data, and $t_b=54137$ days is the time break between two kinematic regimes, which is fixed after the flare. We find an initial offset value $a=0''$.8601$\pm$0$''$.0001, which equals 75.69$\pm$0.01 pc. The slope before the break point ($t_b$) is $b_1$ = 9.97$\pm$0.38$\times 10^{-6}$ $''$/day. The slope after the break point is $b_2$ = 2.00$\pm$0.04$\times 10^{-5}$ $''$/day. The measured speed of HST-1 before $t_b$ is $1.04\pm0.04\,c$ and after $t_b$ is $2.1\pm0.05\,c$. 

\section{Discussion} 
\label{sec:discussion} 

HST-1 is a prominent and complex structure that extends up to a few pc in the downstream region of the jet in M87 \citep[see][for a recent review]{Hada24}. HST-1 has shown significant variability in the X-ray and optical wavelengths. A strong polarization detected at optical frequencies by Hubble robustly indicates that this segment of the emission continuum is synchrotron in origin \citep[e.g.,][]{Perlman11}. It is widely believed that the short-timescale variability seen in the Chandra data for HST-1 is likewise consistent with the X-ray emission being primarily produced by synchrotron processes. A clear correlation exists between the optical and X-ray emission from HST-1. To investigate this correlation, we conducted a comparative imaging analysis of {\it Chandra} and {\it Hubble} observations, focusing on single and multiple emission regions, and the superluminal motion of HST-1.

\subsection{The Flux/Offset Correlation Modeling}

In our previous {\it Chandra} study \citep{Thimmappa24}, we found a correlation between the flux and offset of HST-1 during a bright flare. We developed a toy model involving multiple unresolved emission regions that explains this correlation, and in this work, we apply the same model to data from {\it Hubble}. We collect 245 {\it Hubble}/ACS observations and 120 {\it Hubble}/WFC3 observations covering the period 2002 to 2022, with a total exposure time of 345 ksec. Using these data, we measure the flux of the jet knot HST-1 and its offset from the core.

We apply the toy model (Equations \ref{eqn:flux_HST-1} and \ref{eqn:offset_HST-1}) to the {\it Hubble} data, fitting individual filters to investigate the presence of a single emission component with a fixed working surface velocity of $\sim$6.4 c. Among all filters, F250W provides the best fit, but not a good fit. The distance of the single emission region from the baseline position is still $d<0''.02$ in all bands of {\it Hubble} data. Subsequently, we perform a simultaneous fit of the flux-offset model to the X-ray and {\it Hubble} (UV, optical, and IR) data to investigate the presence of multiple emission regions within HST-1 (see in Appendix~\ref{sec:Chandra_vs_Hubble}). Our fitting results show that the UV/optical and near-IR emission possibly originates from the baseline position of HST-1, while the X-ray emission originates downstream. The measured values suggest that all bands, including UV, optical, and near-IR, remain spatially co-located, with no significant separation among sub-flares, but X-ray emission regions are located elsewhere.

The above-summarized findings point to a general picture in which the observed X-ray and UV/optical emission components originate at different locations along the outflow within the HST-1 complex, that is, at distinct working surfaces. Moreover, they appear to correspond to different energy distributions of the radiating electrons, indicating differences in the particle-acceleration processes operating at the two sites. More specifically, the absence of X-ray emission at the position of the optical knot could be explained if the maximum electron energies available within this steady, baseline region of HST-1 are lower, while those in the X-ray-emitting sites further downstream are higher, enabling synchrotron photons to reach keV energies. Finally, within these X-ray emission zones, the electron energy spectrum must be relatively narrow so as not to outshine, at optical frequencies, the baseline 
optical emission of the HST-1 knot.

Regarding this last point, we note that for a spectral index $\alpha$ characterizing the synchrotron continuum between optical ($\sim$4.6 eV photons) and X-ray ($\sim$3 keV photons) energies, the ratio of the corresponding energy flux densities $\varepsilon F_{\varepsilon}$ (e.g., in erg\,s$^{-1}$\,cm$^{-2}$) is
\begin{equation}
 \frac{[\varepsilon F_{\varepsilon}]_{\rm opt}}{[\varepsilon F_{\varepsilon}]_{\rm X}} 
 \sim [\frac{\varepsilon_{\rm opt}}{\varepsilon_{\rm X}}]^{(1-\alpha)} \sim 0.0015^{(1-\alpha)} 
 \label{eqn:synchrotron}
\end{equation}

This ratio is $\sim 0.04$ for a typical synchrotron index of $\alpha = 0.5$, and can be as low as $\sim 2\times10^{-4}$ for the flattest possible synchrotron spectrum, $\alpha = -1/3$. The maximum flux of HST-1 increases around 2005/2006, and both the X-ray and optical quantities are measured as energy flux densities. For synchrotron emission, the expected UV flux density from the X-ray emitting region is $\sim 0.25\times10^{-12}$\,erg\,s$^{-1}$\,cm$^{-2}$ for $\alpha=0.5$, and $\sim 1.07\times10^{-15}$\,erg\,s$^{-1}$\,cm$^{-2}$ for $\alpha=-1/3$. From equation \ref{eqn:synchrotron}, the upper limit on the F250W flux from the X-ray emitting region is $\sim$4\% of the observed baseline emission. In principle, this could be detectable, but it could also plausibly be hidden behind the brighter baseline region. Ignoring the baseline emission, a nondetectable UV flux enhancement would imply an upper limit of $\alpha \lesssim 0.043$, assuming a flux sensitivity limit $4.75\times10^{-18}$\,erg\,s$^{-1}$\,cm$^{-2}$\AA$^{-1}$ \citep{Calamida22}. In short, optical/UV emission from the X-ray emitter need not be detectable.

Previously, \cite{Biretta99} reported two emission regions within HST-1 using the {\it Hubble}/FOC instrument with a high angular resolution, sampling to a finer scale of $0''.00478$/pixel. In comparison, our analysis is based on Hubble/ACS and WFC3/UVIS observations. These instruments have spatial resolutions approximately 2–4 times lower than the FOC, which limits our ability to resolve similarly fine substructures. Indeed, within the $\sim0''.05$ spatial resolution limit of {\it Hubble}'s ACS/WFC3 data, we find no apparent evidence for multiple optical components in HST-1. Furthermore, we do not detect any features analogous to the compact structures revealed by VLBI at $\sim0.5-4$ mas resolution \citep{Cheung07, Chang10, Giroletti12}.

We interpret the observed optical and X-ray characteristics, both in the temporal and spatial domains, as signatures of a recollimation shock structure developing around the position of HST-1. Such a structure forms when an overpressured, supersonic jet that initially expands freely is subsequently forced back toward its axis by the surrounding medium, compressing the flow and enhancing particle acceleration and magnetic-field compression, particularly around the ``nozzle'' that emerges within the outflow.

The idea that HST-1 marks a reconfinement region of the M87 jet has a long history. \cite{Stawarz06} first emphasized that the extreme multiwavelength flare of HST-1 and its distance from the core could be explained if the feature corresponds to a recollimation nozzle within a relativistic hydrodynamical jet, for plausible AGN and ambient-medium parameters of the M87 system. Indeed, as shown by \cite{Asada12} and \cite{Nakamura13}, the M87 jet morphology transitions from a parabolic to a conical streamline near HST-1, consistent with the formation of a stationary recollimation shock. It is interesting to note that comparable jet features, though generally observable only at lower angular resolution and typically only at radio frequencies, have been identified in the radio-loud narrow-line Seyfert 1 galaxy 1H 0323+342 \citep{Hada18}, the blazar BL Lacertae \citep{Cohen14}, and the radio galaxy 3C 120 \citep{Agudo12}.

Our results indicate that the optical emission of HST-1 is dominated by a single region that is co-spatial with a steady, baseline shock, rather than by individual downstream-moving components. The optical offsets measured from the Hubble data exhibit a smooth, slowly varying trend over two decades, likely tracing the evolution of the recollimation-shock nozzle driven by a gradual modulation in the jet power or in the pressure of the confining ambient medium. Superimposed on this are smaller-scale radio knots traced by VLBI, as well as multiple compact and variable emission regions that dominated the X-ray flux during the 2005/2006 outburst, as discussed by \cite{Thimmappa24}.

In a purely hydrodynamic picture, a related phenomenon --- superluminal plasma blobs passing through a reconfinement nozzle --- was studied by \cite{Komissarov97}, who showed that a fast-moving blob interacting with a hydrodynamic reconfinement shock can brighten dramatically upon encountering the nozzle but then fades rapidly once it passes the reflection point. This rapid downstream fading, however, poses a challenge for explaining the prolonged, large-amplitude variability of HST-1 within a purely hydrodynamic framework. This naturally points instead toward a magnetohydrodynamic (MHD) scenario in which the jet magnetic field is dynamically important, possibly even dominant. In such a strongly magnetized regime, as shown by \cite{Levinson16}, a change in the external pressure profile naturally produces a focusing nozzle followed by a sequence of oscillatory recollimation shocks (``nodes''). The apparent `downstream--then--upstream' motion of the X-ray peak of HST-1, correlated with the X-ray flux and contrasted with the essentially fixed location of the optical flare at the upstream edge of the knot, can possibly be understood in this scenario as the signature of transient internal shocks or fast-mode compressions forming downstream of the nozzle and subsequently fading, causing the flux-weighted X-ray centroid to shift outward and then return.

Furthermore, recent {\it JWST} observations (2024) measured the baseline position of HST-1 at $0''.950\pm0''.05$ from the core, and an additional emission region separated by $\sim0''.15\pm0''.02$ downstream within HST-1 in the infrared band \citep{Roder25}. These IR knots are located at the baseline position of HST-1 compared with the optical emission, but are offset by $\sim0''.1$ relative to the radio emission. The inability to reconcile the {\it Chandra} and {\it Hubble} datasets with a single emission region strongly suggests different jet characteristics at different wavelengths, underscoring the critical importance of high-resolution, multi-wavelength observations in resolving the radiation physics of relativistic jets.

\subsection{The Sub and Superluminal Motion of HST-1}

Additionally, we perform proper motion measurements of the jet using the {\it Hubble} data. We find a gradual increase in the offset over time, with the measured bulk speed of 1.04$\pm$0.04\,c until 2006. This speed corresponds to a particular bright substructure within HST-1 that is moving relatively slowly. From 2009 to 2022, we measured superluminal motion of 2.4$\pm$0.5\,c in HST-1. A piecewise regression model, however, reveals a clear transition from v=1.04$\pm$0.04 c before 2005 to 2.1$\pm$0.05 c thereafter. Our measurements (Sections \ref{sec:velocity_calculation1}-\ref{sec:velocity_calculation3}) of HST-1 confirm a clear transition from slower to faster motion across a temporal break. The components of HST-1 move at different speeds in different locations, initially moving slowly before the temporal break and then faster afterward.

This temporal acceleration of HST-1 is crucial for interpreting the spatial acceleration along the jet. By observing changes in the velocity of HST-1 over time and distance, we build a connection between temporal and spatial acceleration. Comparable spatial accelerations are observed in other radio-loud AGN jets. For example, the jet of Cygnus A exhibits apparent spatial acceleration within its inner regions, with components around $\sim0.54$ c/pc along its jet before they slow near the terminal hotspot \citep{Krichbaum98, Boccardi16}. Similarly, VLBA studies of NGC 315 demonstrate that the spatial acceleration of the jet knots is $\sim0.47$ c/pc \citep{Cotton99, Canvin05}. In M87 itself, \cite{Meyer17} note that the jet exhibits continued spatial acceleration up to $\sim$100 pc from the core. For comparison, we see an apparent acceleration of HST-1 from a speed of 1.04$\pm$0.04 c in 2005 to 2.1$\pm$0.05 c in 2022. 

During the period between MJD 54137 and 55763 (i.e., 2006 and 2011), our model indicates that the speed of HST-1 changes by $\Delta v$= 1.06 c, while it moves a distance $\Delta r$ = 3.68 pc. This corresponds to a spatial acceleration ($\Delta v/\Delta r$) of 0.288 c/pc. The spatial acceleration of HST-1 is lower than the values reported for Cygnus A (0.54 c/pc) and NGC 315 (0.47 c/pc). However, the presence of high acceleration in other well-studied jets supports the plausibility of our measurement and suggests that the observed acceleration is physically reasonable within the jets.

The velocities we measure with respect to time align well with previous optical and radio results, supporting long-term bulk superluminal motion in the jet of M87. For example, the previously measured values of HST-1 is $0.61\pm0.31$ c from 2003 to 2006 \cite[VLBI,][]{Chang10}, and $1.23\pm0.91$ c from 2003 to 2007 \cite[VLA,][]{Chen11a}. The VLBA $\lambda20$ cm observations from 2005 to 2006 reported apparent velocities of 0.49$\pm$0.39 c for the HST-1\,c2 component and 1.14$\pm$0.14 c for HST-1 \,d component \citep{Cheung07}. The VLBI studies \citep[e.g.,][]{Giroletti12} have modeled the internal structure of HST-1 and reported three distinct components: two moving at $\sim$ 4.1 c and one at 6.4 c over a $\sim$5-year period.  Similarly, \cite{Asada14} reported the superluminal motion of HST-1 is 2.5 c and 3.5 c from 2007 to 2009. From the MOJAVE program, \cite{Lister13} measured apparent speeds in the range of $\sim$1.4 - 4.6 c in parsec-scale regions of the jet. The coexistence of slow and fast components within HST-1 has been interpreted in various ways. Slower speeds observed in lower-frequency VLBI observations have been attributed to standing shocks \citep{Walker08, Kovalev07}. This velocity stratification is also consistent with a spine-sheath jet structure, where a sub-relativistic sheath surrounds a relativistic spine, and faster-moving knots are detected in optical/X-ray observations \citep{Nakamura13, Mertens16}.

The velocity variation in the substructures of HST-1 may also reflect differences in the emission regions across wavelengths. In particular, for X-rays, higher speeds than the optical/UV. {\it Chandra} observations yield apparent velocities of $6.3\pm0.4$ c \citep{Snios19} and $6.6\pm0.9$ c \citep{Thimmappa24}, associated with flaring substructures. The higher velocities inferred from internal substructure further support the interpretation of localized shocks within HST-1 as the origin of high-energy flaring activity. The optical emission regions might not trace the higher velocity material modeled in the X-ray band. The acceleration we observe follows the emergence of multiple flares and compact substructures in X-ray imaging, supporting an interpretation involving internal shocks within the jet. 

Together, these results highlight the importance of multiwavelength studies of jet kinematics. HST-1 serves as a key observational tool for understanding the relativistic dynamics of the jet in M87 and potentially for jets in other AGN. Our study provides a crucial link between inner jet acceleration, localized flaring behavior, and the long-term propagation of superluminal components. These results provide important insights into the energy dissipation and radiative processes governing jet flares, advancing our understanding of jet physics in active galactic nuclei.

This research used data from the \textit{Hubble} Data Archive. This work was supported by NASA award 80NSSC20K0645 (R.T., J.N.). The authors thank the anonymous referee for the valuable comments on the manuscript.

Facility: HST (ACS, WFC3) - {\it Hubble} Space Telescope satellite.

\textit{Software:} 
Numpy \citep{van_der_Walt11}, Astropy \citep{Astropy_Collaboration13, Price-Whelan18}, Scipy \citep{Virtanen20}, Matplotlib \citep{Hunter07}, Pandas \citep{McKinney10}, CIAO v4.17 \citep{Fruscione06}, Sherpa v4.17.0 \citep{Freeman01}, DS9 \citep{Joye03}

\bibliography{ms}

\appendix

\section{Joint Fitting with {\it Chandra} and {\it Hubble}}
\label{sec:Chandra_vs_Hubble}

Our {\it Chandra} study \citep{Thimmappa24} revealed multiple unresolved emission regions for the flux-offset correlation of HST-1 during a flare event. Following up on this study, we perform a simultaneous fit for the X-ray and UV/optical data using the same model functions (i.e., Eqn \ref{eqn:flux_HST-1} and \ref{eqn:offset_HST-1})
to determine whether the {\it Hubble} data is consistent with the additional emission regions observed by {\it Chandra}. As we see from Figure~\ref{fig:Offset_ACS_vs_ACIS}, which shows that the offset is not clearly correlated to the flux of HST-1. However, to quantify additional emission regions in more detail, we apply our toy model to Hubble data along with X-ray data by simultaneously fitting. Simultaneous fit helps us to understand the origin of X-ray emission and UV bands and the relationship between flare events. This approach is important for studying relativistic jets and understanding the inner working surface mechanism.

We perform a joint fit to reconcile the data at different wavelengths, given the discrepancy between the {\it Chandra} data (which we modeled with multiple emission regions) and the {\it Hubble} data (which do not appear to need them). For the simultaneous fit, we added two additional emission components to the flux/offset model to find the locations of flaring regions ($d_1$, $d_2$, and $d_3$) from the X-ray and {\it Hubble} data. In the fit, the initial offset $D$ and the additional flare distances $d_1$, $d_2$, and $d_3$ are linked across all filters to the {\it Chandra} data. Previously, for {\it Chandra} data, we found the initial offset (baseline position) from the core was $D=0''.8536\pm0''.0012$, and the extra emission regions were located at distances from the baseline position of $d_1=0''.2486\pm0''.0074$, $d_2=0''.1918\pm0''.0037$, and $d_3=0''.0811\pm0''.0037$. From the joint fit, we find distances of $d_1=0''.0531\pm0''.0020, d_2=0''.3193\pm0''.01094$, and $d_3=0''.0694\pm0''.0023$ respectively. These measured distances are much closer to the baseline of HST-1 and are inconsistent with our previous {\it Chandra} results. More importantly, our model does not fit the data well: the reduced $\chi^{2}$ of the model is 13.064. The model to the X-ray and optical data provides a poor fit for the same regions inside HST-1.

\begin{figure*}
    	\centering 
\includegraphics[width=0.72\textwidth]{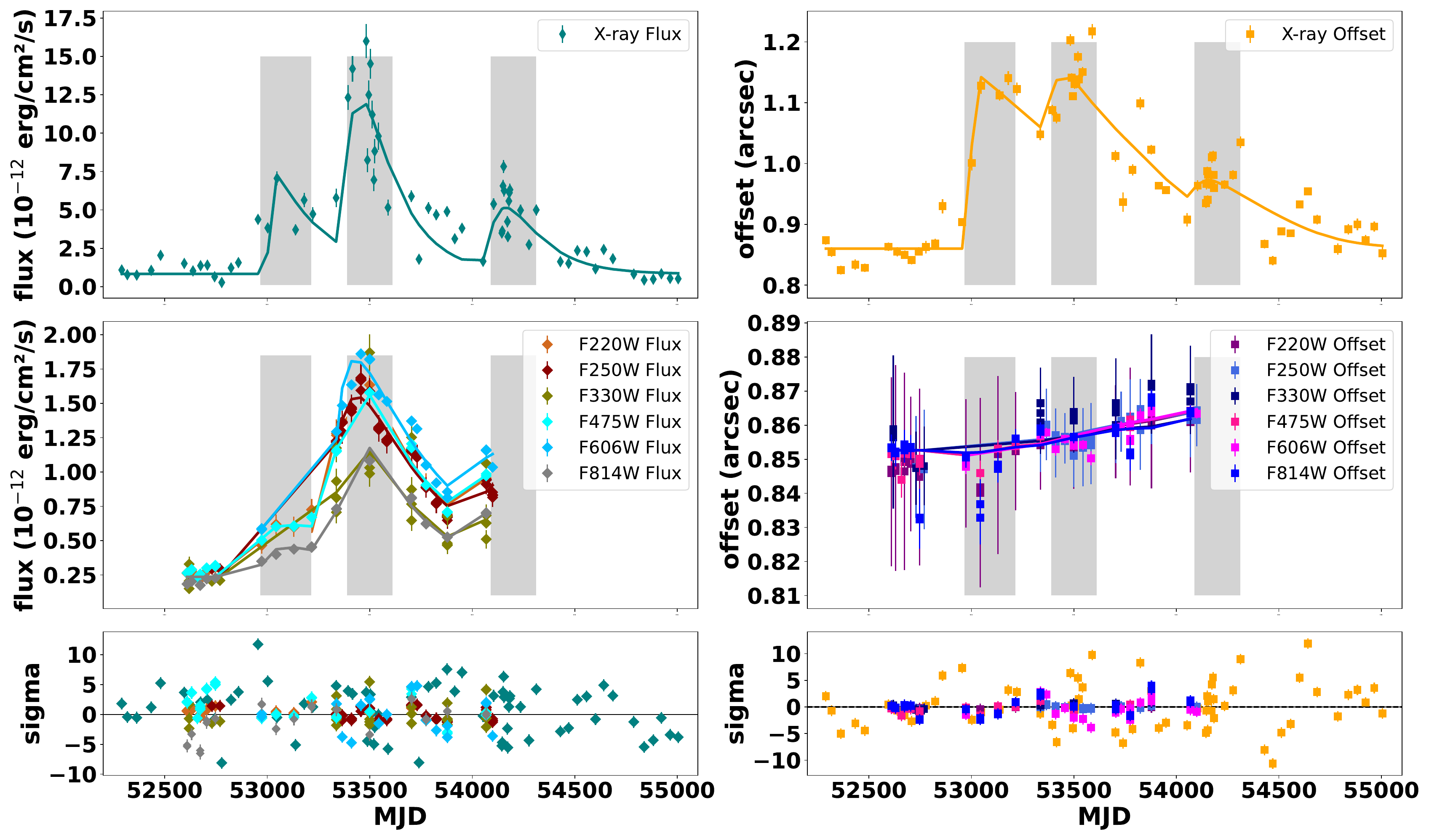}
	\caption{Simultaneous fits to the lightcurve and offset of X-ray and {\it Hubble} data. Mainflare, subflare 1, and 2 are shown in gray color.}
   \label{fig:X-ray_vs_Hubble}
\end{figure*}

Next, the distances $d_1$, $d_2$, and $d_3$ are set free in the {\it Hubble} data, allowing them to vary independently of the X-ray. The working surface velocity ($V_{ws}$) is also allowed to vary for {\it Hubble} data. The resulting parameters are given in Table \ref{tab:1_simultaneous_fitting_Chandra_Hubble}, and the corresponding model fit is shown in Figure \ref{fig:X-ray_vs_Hubble}.

The initial offset $D$ is 0.$''$8526$\pm$0.$''$0006, equal to 75.03$\pm$0.05 pc. For the main flare, the {\it Hubble} data exhibits a significantly lower amplitude ($0.49\pm0.01\times 10^{-12}$ erg/cm$^2$/s) than the X-ray data (8.6$\pm$0.4$ \times 10^{-12}$ erg/cm$^2$/s), which shows that the X-ray flux dominates during the peak phase. The start time is earlier in the {\it Hubble} bands (52930$\pm$5 days) than in X-rays (53000$\pm$5 days). The fall and rise times are longer in UV/optical (430$\pm$1 days and 75$\pm$1 days) than in X-rays (240$\pm$1 days and 5$\pm$1 days). This change between rise and fall times shows that the optical emitting electrons cool more slowly than the X-ray emitting electrons. For subflare 1 and 2, the amplitudes of X-ray data are $48\pm2\times10^{-12}$ erg/cm$^2$/s and 
$30\pm1\times 10^{-12}$ erg/cm$^2$/s, respectively. The fall times are shorter than the main flare, about $150\pm$5 days for both, and the rise time is $100\pm$1 and 150$\pm$1 days. In {\it Hubble} observations, the rise time of subflare 1 is 90$\pm$5 days, which is higher than that of subflare 2. The rise time of subflare 2 is higher than that of both mainflare and subflare 1 in X-rays, which suggests a longer cooling time of this region. The additional emission region distances $d_1$, $d_2$, and $d_3$ from X-ray data are resolved and similar to our previous {\it Chandra} \citep{Thimmappa24} results at 0$''$.253$\pm$0$''$.007, 0$''$.188$\pm$0$''$.004, 0$''$.080$\pm$0$''$.003, respectively. However, the same parameters derived from {\it Hubble} are negligible or close to zero (-0$''$.003$\pm$0$''$.001, 0$''$.0004$\pm$0$''$.0011, 0$''$.011$\pm$0$''$.002). These results indicate that the X-ray and optical emissions could arise from different locations, which implies different particle acceleration processes in the different shock regions.

\begin{deluxetable*}{ccccc c}
\tabletypesize{\normalsize}
\tablecaption{The resulting parameters of the simultaneous fitting of {\it Chandra} X-ray and {\it Hubble} ACS data.}
\label{tab:1_simultaneous_fitting_Chandra_Hubble}

\tablehead{
\colhead{Flares} & \colhead{Parameters} & \colhead{Symbol} & \colhead{X-ray} & \colhead{Hubble} & \colhead{Units}
}
\startdata
Baseline    & Flux          & F$_{o}$     & 0.84$\pm$0.03 & 0.236$\pm$0.002 & 10$^{-12}$ erg cm$^{-2}$ s$^{-1}$ \\
            & Offset        & D           & 0.8526$\pm$0.0006 & linked & arcsec \\
            \hline
Main flare  & Amplitude     & A$_1$       & 8.6$\pm$0.4 & 0.49$\pm$0.01 & 10$^{-12}$ erg cm$^{-2}$ s$^{-1}$ \\
            & Start Time    & t$_{o1}$    & 53000$\pm$5 & 52930$\pm$5 & days \\
            & Fall Time     & $\tau_{1}$  & 240$\pm$1 & 430$\pm$1 & days \\
            & Rise Time     & $\tau_{2}$  & 5$\pm$1 & 75$\pm$1 & days \\
            & Distance      & d$_1$       & 0.253$\pm$0.007 & $-$0.003$\pm$0.001 & arcsec \\
\hline
Subflare 1  & Amplitude     & A$_2$       & 49.7$\pm$2.1 & 2.91$\pm$0.07 & 10$^{-12}$ erg cm$^{-2}$ s$^{-1}$ \\
            & Start Time    & t$_{o2}$    & 53340$\pm$5 & 53300$\pm$2 & days \\
            & Fall Time     & $\tau_{1}$  & 150$\pm$5 & 240$\pm$5 & days \\
            & Rise Time     & $\tau_{2}$  & 100$\pm$1 & 90$\pm$5 & days \\
            & Distance      & d$_2$       & 0.188$\pm$0.004 & 0.0004$\pm$0.0011 & arcsec \\
\hline
Subflare 2  & Amplitude     & A$_3$       & 30.0$\pm$1.3 & 0.69$\pm$0.02 & 10$^{-12}$ erg cm$^{-2}$ s$^{-1}$ \\
            & Start Time    & t$_{o3}$    & 54020$\pm$1 & 53980$\pm$5 & days \\
            & Fall Time     & $\tau_{1}$  & 150$\pm$5 & 50$\pm$1 & days \\
            & Rise Time     & $\tau_{2}$  & 150$\pm$1 & 70$\pm$6 & days \\
            & Distance      & d$_3$       & 0.080$\pm$0.003 & 0.011$\pm$0.002 & arcsec \\
            \hline
            & Velocity      & $V_{ws}$    & 6.4 (fixed) & 1.9$\pm$0.2 & c \\
\hline
\enddata

\tablecomments{F$_0$ is the constant flux of baseline position of HST-1, $D$ is the initial offset of HST-1 from the core, $V_{ws}$ is the working surface velocity, which is fixed for X-ray data, but allowed to vary for Hubble data for all flares (mainflare, subflare 1, and subflare 2) and linked across all filters. All parameters corresponding to the {\it Hubble} ACS filters (F220W, F250W, F330W, F475W, F606W, and F814W) are linked accordingly.}
\end{deluxetable*} 

\end{document}